\documentclass[aps,prl,twocolumn,showpacs,groupedaddress]{revtex4}
\usepackage{graphicx}
\begin{document}

\title{Thermal entanglement of Bosonic atoms in an optical lattices with nonlinear couplings}
\author{L. Zhou, X. X. Yi, H. S. Song and Y. Q. Quo}
\affiliation{Department of Physics, Dalian University of
Technology, Dalian 116024, China}
\date{\today}

\begin{abstract}
The thermal entanglement of two spin-1 atoms with nonlinear
couplings in an optical lattices is investigated in this paper. It
is found that the nonlinear couplings favor the thermal
entanglement creating. The dependence of the thermal entanglement
in this system on the linear coupling, the nonlinear coupling, the
magnetic field and temperature is also presented. The results show
that the nonlinear couplings really change the feature of the
thermal entanglement in the system, increasing the nonlinear
coupling constant increases the critical magnetic field and the
threshold temperature.
\end{abstract}

\pacs{ 03.67.Mn, 03.67.-a, 32.80.Pj} \maketitle

Entanglement as a valuable resource for quantum information
processing (QIP) \cite{bennett, nielsen1, plenio1} has attracted a
lot of attention in recent years, from both experimental and
theoretical studies \cite{raimond}. Since the entanglement is
fragile, the problem of how to create stable entanglement remains
a main focus of recent studies in the field of quantum information
processing. The thermal entanglement, which differs from the other
kind of entanglement by its advantages of stability, requires
neither measurement nor controlled  switching of interactions in
the preparation process, hence the thermal entanglement in various
systems is an attractive topic and worth intensively studying.

The system of atoms in optical lattices is  among the promising
candidates for quantum information processing. It may take the
advantage of  the technology  used in atom optics and laser
cooling based on the optical manipulation of atoms \cite {birkl}.
Besides, it also holds the merit of eventual possibility to scale,
parallelize and miniaturize the device in QIP.

The thermal entanglement has been extensively  studied for various
systems including isotropic Heisenberg chain \cite{arnesen,
oconnor, iso}, anisotropic Hensenberg chain \cite{aniso}, Ising
model in an arbitrarily directed magnetic field \cite {ising}, and
cavity-QED \cite{cqed} since the seminal works by Arnesen {\it et
al.} \cite {arnesen} and Nielsen  \cite{nielsen2}. Based on the
tools developed within the context of quantum information theory,
the relaxation of a quantum system towards the thermal equilibrium
is investigated \cite {scarani}, which provides us a different
mechanism to model  a system arriving at the thermal entangled
states.  For a specific Heisenberg chain in condensed matter
physics, the only ranging variables are the magnetic field and the
temperature \cite {arnesen}, in cavity-QED system, the exchange
constant (the linear coupling in our case) is adjustable in
addition to the temperature and magnetic field. The development of
laser cooling and trapping provides us more ways to control the
atoms in traps. Indeed, we can manipulate the atom-atom coupling
constants and the atom number in each lattice well with a very
well accuracy \cite{greiner,yip}.

In this paper, we study the thermal entanglement in optical
lattice with nonlinear couplings. We calculate the thermal
entanglement as a function of the nonlinear coupling constant,
linear coupling constant, the temperature  as well as the external
magnetic field. We will confine ourself in this paper to the case
of $K<0$ and $J<0$ that is relevant to the recent experiment
conducted on $^{23}Na$ atoms. As we will show you later on, there
is no thermal entanglement in the regime of $K>J$ when $J<0$
similar to the results  for the isotropic Heisenberg model \cite
{arnesen}. Our studies  also show that the critical magnetic field
and the threshold temperature is obviously increased by the
presence of the nonlinear couplings.

Our system consists of two wells in the optical lattice with one
spin-1 atom in each well. The lattice may be formed by three
orthogonal laser beam, and we may use an effective Hamiltonian of
the Bose-Hubbard form \cite{jaksch} to describe the system. The
atoms in the Mott regime make sure that each well contains only
one atom. For finite but small hopping term $t$, we can expand the
Hamiltonian into powers of $t$ and get \cite{yip},
\begin{equation}
H=\epsilon +J(\vec{S}_1\cdot \vec{S}_2)+K(\vec{S}_1\cdot
\vec{S}_2)^2, \label{Ha1}
\end{equation}
where $J=-\frac{2t^2}{U_2},K=-\frac{2t^2}{3U_2}-\frac{4t^2}{U_0}$
with $t$   the hopping matrix elements, and  $\epsilon =J-K$.
$U_s$ ($s=0,2)$ represents the Hubbard repulsion potential with
total spin $s$, a potential  $U_s$ with $s=1$ is not allowed due
to the identity of the bosons with one orbital state per well,
$\vec{S}_i(i=1,2)$ denote the spin vector
$\vec{S}_i=(S_{ix},S_{iy,}S_{iz})$. This Hamiltonian differs from
the usual Heisenberg model by the nonlinear couplings.  Since term
$\epsilon$ contains no interaction, we can ignore it in the
following discussions and it would not change the thermal
entanglement. In the presence of external magnetic field, the
Hamiltonian Eq. (\ref{Ha1}) becomes
\begin{equation}
H=J(\vec{S}_1\cdot \vec{S}_2)+K(\vec{S}_1\cdot
\vec{S}_2)^2+B(S_{1z}+S_{2Z}),\label{Ha2}
\end{equation}
where the magnetic field $\vec{B}$ along the z-direction is
assumed. When the total spin for each site $S_i=1 (i=1,2)$, its
components take the form
\begin{widetext}
\begin{equation}
S_{ix}=\frac 1{\sqrt{2}}\left(
\begin{array}{ccc}
0 & 1 & 0 \\
1 & 0 & 1 \\
0 & 1 & 0
\end{array}
\right) ,S_{iy}=\frac 1{\sqrt{2}}\left(
\begin{array}{ccc}
0 & -i & 0 \\
i & 0 & -i \\
0 & i & 0
\end{array}
\right) ,S_{iz}=\left(
\begin{array}{ccc}
1 & 0 & 0 \\
0 & 0 & 0 \\
0 & 0 & -1
\end{array}
\right) .
\end{equation}
\end{widetext}
The ground state of Hamiltonian Eq.(\ref{Ha2}) is expected to be
the dimer phases, as the most recently study \cite{yip} conclude,
this is quite different from the Heisenberg chain without
nonlinear couplings. And the nonlinear couplings  would make the
thermal entanglement different from that in the usual Heisenberg
model, too, as you will see.

To get the thermal entanglement, we first present the eigenvalues
and the corresponding eigenstates of the Hamiltonian
Eq.(\ref{Ha2}),
\begin{eqnarray}
E_1 &=&K+J-B;|\Psi _1\rangle =\frac 1{\sqrt{2}}(|0,-1\rangle
+|-1,0\rangle ),\nonumber
\\
E_2 &=&K+J+B;|\Psi _2\rangle =\frac 1{\sqrt{2}}(|1,0\rangle
+|0,1\rangle ),\nonumber
\\
E_3 &=&K-J+B;|\Psi _3\rangle =\frac 1{\sqrt{2}}(-|1,0\rangle
+|0,1\rangle ),\nonumber
\\
E_4 &=&K-J-B;|\Psi _4\rangle =\frac 1{\sqrt{2}}(-|0,-1\rangle
+|-1,0\rangle
), \nonumber \\
E_5 &=&K+J;|\Psi _5\rangle =\frac 1{\sqrt{6}}(|1,-1\rangle
+|-1,1\rangle
+2|0,0\rangle ),\nonumber  \\
E_6 &=&K-J;|\Psi _6\rangle =\frac 1{\sqrt{2}}(|1,-1\rangle
-|-1,1\rangle ),\nonumber
\\
E_7 &=&K+J+2B;|\Psi _7\rangle =|1,1\rangle , \nonumber \\
E_8 &=&K+J-2B;|\Psi _8\rangle =|-1,-1\rangle , \nonumber \\
E_9 &=&4K-2J;|\Psi _9\rangle =\frac 1{\sqrt{3}}(|1,-1\rangle
+|-1,1\rangle -|0,0\rangle ).\nonumber \\
\label{eigen}
\end{eqnarray}
The state of the above system at thermal equilibrium is $  \rho
=Z^{-1}\exp (-\beta H)$, where $Z=Tr[\exp (-\beta H]$ is the
partition function and $\beta =1$ $/k_BT$ ($k_B$ is Boltzmann's
constant. For simplicity we will set $k_B=1$ hereafter). In terms
of the eigenstates and the corresponding eigenvalues, the state of
the system can be expressed as
\begin{equation}
\rho =\frac 1Z\sum_{i=1}^9e^{-\beta E_i}|\Psi _i\rangle \langle
\Psi _i|
\end{equation}
with the partition function
\begin{widetext}
\begin{equation}
Z=2e^{-\beta K}\cosh(\beta B)(1+2\cosh(\beta
J))+2e^{-\beta(K+J)}\cosh(2\beta B)+e^{-\beta(4K-2J)}.
\end{equation}
\end{widetext}
We will choose the negativity as the entanglement measure
\cite{vidal},
\begin{equation}
N(\rho )=\frac{\left\| \rho ^{T_A}\right\| -1}2,
\end{equation}
where $\left\| \rho ^{T_A}\right\| $ denotes the trace norm of the
partial transpose $\rho ^{T_A}$. The negativity $N(\rho )$ is
equivalent to the absolute value of the sum of the negative
eigenvalues of $\rho ^{T_A}$\cite {horodecki}. Although the
negativity lacks a direct physical interpretation, it bounds two
relevant quantities in quantum information processing--the channel
capacity and the distillable entanglement. As the negativity is a
computable measure of entanglement for bipartite system with any
dimension, we here choose it to measure the thermal entanglement.

We have performed extensive numerical calculations for the
entanglement measure, some selected results are presented in
figures from 1 to 3.
\begin{figure}
\includegraphics*[width=0.95\columnwidth,
height=1.2\columnwidth]{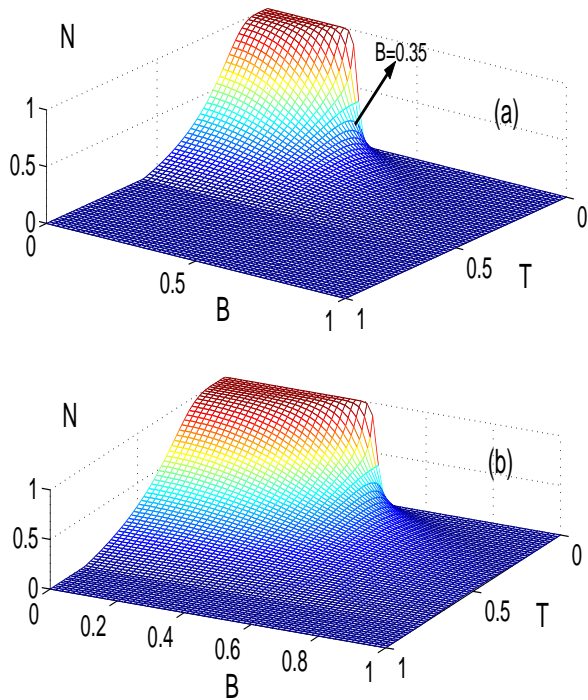} \caption{(Color on line) The
thermal entanglement {\it vs.} the external magnetic field $B$ and
temperature. We choose the negativity as the entanglement measure,
$B$ and $T$ are plotted in units of the Boltzmann's constant
$k_B$. At $T=0$, $N$ has a sharp transition from 1 to 0 as $B$
crosses the critical value of $B=3/2(J-K)$. The $B=0.35$ line
pointed  out in the figure shows that for certain value of $B$ it
is possible to increase $N$ by increasing temperature $T$. Figure
1-(a), (b) are for different $K$, (a) $K=-0.6$ and (b) $K=-0.7$. }
\label{fig1}
\end{figure}
Figure 1 shows the plot of the negativity as a function of the
magnetic field $B$ and temperature $T$. For $B=0$, the state
$|\Psi _9\rangle $ is the ground state \cite{exp}, and the others
in Eq. (\ref{eigen})  are excite states. In this case, the maximal
entanglement is at $T=0$ and it decreases with $T$ due to mixing
of the excited states with the ground state. For a higher value of
$B$ greater than $B_c=3/2(J-K)$, $|\Psi _8\rangle $ becomes the
ground state. In that case, there is no entanglement at $T=0$, but
we may increase the entanglement by increasing $T$, that is to
bring entangled eigenstates such as $|\Psi_i\rangle$
$(i=1,2,3,4,5,6,9)$ into mixing with the ground state. It is
interesting to note that the critical field $B_c$ depends on both
the linear coupling $J$ and the nonlinear coupling $K$. With
$K=0$, $B_c$ gives rise to $3/2 J$ that means no entanglement in
this case (i.e. for $J<0$) at any temperature and with any values
of magnetic field. We would like to address that $J$, $K$ and $B$
together determine the ground state properties of the system
instead of $J$ and $B$ in the isotropic Heisenberg model. This is
a quite different feature from the previous studies. For example,
we may choose $K$, $J$ and $B$ such that $|\Psi_1\rangle$ is the
ground state of the system, it is entangled state but not a
maximally entangled one. The critical magnetic field can be
increased by increasing the nonlinear coupling $|K|$, as shown in
figure 1-(b), where the entanglement is plotted as a function of
$T$ and $B$ with the same parameters as in figure 1-(a), but
$K=-0.7$. It is obvious that the threshold temperature above which
the entanglement vanishes has also been increased. More clearly,
this point was shown in figure 2, where we plot the thermal
entanglement in the system as a function of $K$ and $T$.
\begin{figure}
\includegraphics*[width=0.95\columnwidth,
height=1.2\columnwidth]{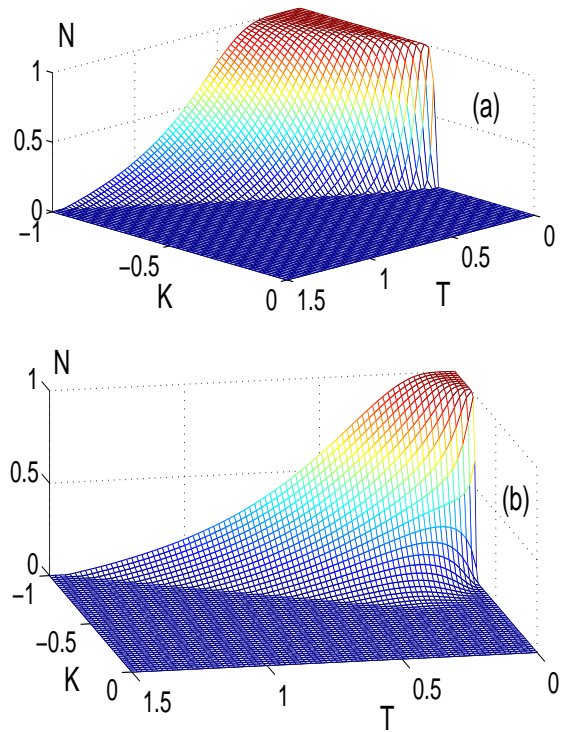} \caption{(Color on line) This
figure depicts the thermal entanglement as a function of the
temperature $T$ and the nonlinear coupling constant $K$ with a
specific $J=-0.4$. The degree of entanglement is measured in terms
of the negativity. Figure 2-(a), (b) are for different $B$, (a)
$B=0$, and (b) $B=0.5$. All parameters are re-scaled in units of
$k_B$. } \label{fig2}
\end{figure}
For $B=0$, the thermal entanglement is not zero only for
$|K|>0.4$, as the figure 2-(a) shows, this indicates that
$|K|>|J|$ is a necessary condition for thermal entanglement to
exist. Further analysis shows that for $B=0$ and with a specific
temperature $T$  the condition for thermal entanglement to exist
is
\begin{equation}
K<J-\frac{T}{3} ln(\frac{5+3e^{2J/T}}{2}). \label{con}
\end{equation}
The condition Eq.(\ref{con}) also holds for $J>0$ and $K>0$, since
we made no constraint on the derivation of Eq.(\ref{con}). On the
other hand, it gives rise to a analytical expression for the
threshold temperature $T_c$ in the case of $B=0$ by solving
$K=J-\frac{T_c}{3} ln(\frac{5+3e^{2J/T_c}}{2})$, it shows again
that $T_c$ depends on $J$ and $K$. In addition to the above
feature,  there is a evidence that the threshold temperature is a
monotonous function of $|K|$ with $B=0$, the larger the nonlinear
coupling constant $|K|$, the larger the threshold temperature.
This point would  be changed when the  magnetic field is present
(Fig. 2-(b)), as you see the threshold temperature is no longer a
monotonous function of $|K|$. We now look at the dependence of the
thermal entanglement on $J$ and $K$, with a fixed $B$ and $T$, the
selected results for this dependence were illustrated in figure 3.
As figure 3 shows, the thermal entanglement may exist only  in the
regime of larger $|K|$, or smaller $|J|$.
\begin{figure}
\includegraphics*[width=0.95\columnwidth,
height=0.6\columnwidth]{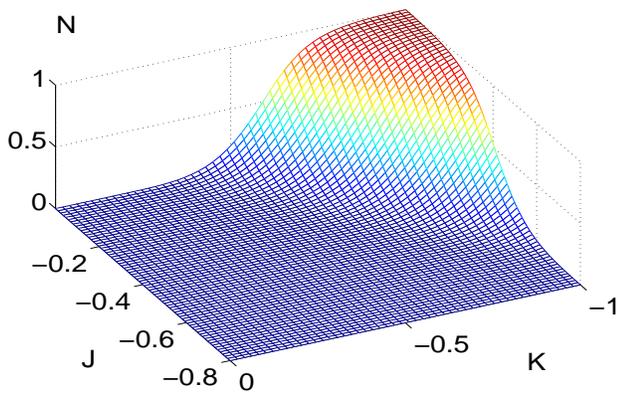} \caption{ (Color on line)
Thermal entanglement {\it vs.} the coupling constants $J$ and $K$.
The selected results are chosen from the regime of $J, K <0$,
which is relevant to atoms $^{23}Na$ in an optical lattice. The
parameters chosen are $B=0.8$ and $T=0.2$. All parameters are
rescaled in units of $k_B$, the Boltzmann's constant. }
\label{fig3}
\end{figure}

Now we discuss the experimental feasibility for observing the
thermal entanglement in the optical lattice. We may choose
$^{23}Na$ trapped in an optical lattice as the system.
Bose-Einstein condensates of unpolarized $^{23}Na$ have already
been achieved by the MIT group \cite{stenger}. The $^{23}Na$ atoms
have hyperfine spin 1, and the interaction among them is
antiferromagnetic  ($t>0$) that is essential for the thermal
entanglement to exist  in isotropic Heisenberg model. There are
two parameters, the hopping matrix $t$ and the Hubbard repulsion
$U_2$ (or $U_0$), we may control separately by adjusting the
strength of the sinusoidal potentials via the intensities of the
laser beams, the coupling constants for the linear interaction $J$
and the nonlinear term are hence adjustable. The two wells in the
optical lattice in question can be isolated by increasing the
barriers connecting to the other wells in the optical lattice.
With these done,  the two $^{23}Na$ atoms in the Mott regime then
end up in a thermal entangled state.

To conclude, we have studied the thermal entanglement in a optical
lattice, the dependence of the thermal entanglement measured by
the negativity on the linear coupling constant $J$, the nonlinear
coupling constant $K$, the external magnetic field and temperature
was presented and discussed. There are two different points in
contract to the isotropic Heisenberg model, one is the
dimensionality and another is the nonlinearity of the couplings.
When the nonlinear coupling is zero, there is no thermal
entanglement in the regime $J>0$, this is similar to that in the
isotropic Heisenberg model in spite of different dimensionality.
The nonlinear coupling really change the feature of the thermal
entanglement, increasing $|K|$ increases the critical magnetic
field and the threshold temperature. The thermal entanglement in
an optical lattice with more than two coupled wells remains
untouched. In future, we will investigate these problems and study
how to map this natural entanglement onto photons and use it as a
resource in QIP.

This work was supported by EYTP of M.O.E, and NSF of China.


\begin{references}
\bibitem{bennett} C. H. Bennett and D. P. DiVincenzo, Nature {\bf
404}, 247(2000).

\bibitem{nielsen1}  M. A. Nielsen and I. L. Chuang, Quantum computation and
quantum information (Cambridge University press, Cambridge, 2000).

\bibitem{plenio1} M. B. Plenio and V. Vedral, Contemp. Phys. {\bf
39},431(1998).

\bibitem{raimond} See, for example, J. M. Raimond {\it et al.,} Rev. Mod.
Phys. {\bf 73}, 565(2001); Vedral, Rev. Mod. Phys. {\bf74},
197(2002).

\bibitem{birkl} G. Birkl {\it et al.,} Optics Comm. {\bf 191}, 67
(2001).

\bibitem{arnesen}  M. C. Arnesen, S. Bose and V. Vedral, Phys. Rev. Lett. {\bf 87},
017901 (2001).

\bibitem{oconnor}  K. M. OConnor and W. K.Wootters, Phys. Rev. A {\bf 63}, 052302
(2001).

\bibitem{iso}  X. Wang, Phys. Rev. A {\bf 66} ,044305 (2002); X. Wang, Phys. Rev. A {\bf 66} , 034302
(2002).

\bibitem{aniso}X. Wang, Phys. Rev. A {\bf 64} , 012313 (2001);G. L. Kamta and A. F. Starace, Phys. Rev. Lett. {\bf 88},
107901(2002);L. Zhou, H. S. Song, Y. Q. Guo and C. Li, Phys. Rev.
A {\bf 68}, 024301 (2003).

\bibitem{ising} D. Gunlycke {\it et al.,} Phys. Rev. A {\bf 64},
042302 (2001).

\bibitem{cqed} S. Mancini, and S. Bose, e-print quant-ph/0111055.

\bibitem{nielsen2} M. A. Nielsen, e-print quant-ph/0011036.

\bibitem{scarani} V. Scarani {\it et al.,} Phys. Rev. Lett. {\bf
88}, 097905 (2002).

\bibitem{greiner} M. Greiner {\it et al.,} Nature {London} {\bf
415}, 39(2002).

\bibitem{yip}  S. K. Yip, Phys. Rev. Lett. {\bf 90}, 250402 (2003).


\bibitem{jaksch}  D. Jaksch, C. Bruder, J.I. Cirac, C. W. Gardiner, and P.
Zoller, Phys. Rev. Lett. {\bf 81},3108 (1998).

\bibitem{vidal}  G. Vidal, R. F. Werner, Phys. Rev. A {\bf 65},
032314 (2002).

\bibitem{horodecki}  K. Zyczkowski, P. Horodecki et al, Phys. Rev. A {\bf 58},
883(1998).

\bibitem{exp} It depends on the parameters $J$ and $K$ chosen. For
an optical lattice with $^{23}Na$ atoms, it is shown that $K<J<0$.

\bibitem{stenger} J. Stenger {\it et al.,} Nature (London) {\bf
396}, 245 (1999).
\end{references}
\end{document}